\documentstyle[epsf]{article}
\def\abstract#1{\vskip 7mm 
        \begin{center}{\large Abstract}\par \smallskip
                \begin{minipage}[c]{12cm}
                        \small #1
                \end{minipage}
        \end{center}
}
\def\title#1{\begin{center}{\Large\bf #1}\end{center}}
\def\author#1{\vskip 5mm \begin{center}{#1}\end{center}}
\def\address#1{\begin{center}{\it #1}\end{center}}
\def\eq{\begin{equation}}
\def\en{\end{equation}}
\def\hbar{(h/2\pi)}
\newcommand{\beqn}{\begin{eqnarray}} 
\newcommand{\eeqn}{\end{eqnarray}} 
\makeatletter
\@ifundefined{lesssim}{\def\lesssim{\mathrel{\mathpalette\vereq<}}}{}
\@ifundefined{gtrsim}{}{}
\def\vereq#1#2{\lower3pt\vbox{\baselineskip1.5pt \lineskip1.5pt
\ialign{$\m@th#1\hfill##\hfil$\crcr#2\crcr\sim\crcr}}}
\makeatother

\begin{document}

\title{%
  Quantum Gravity with Minimal Assumptions
  \smallskip \\
}
\author{%
  Miyuki Nishikawa \footnote{E-mail:nisikawa@hep-th.phys.s.u-tokyo.ac.jp}
}
\address{%
  Department of Physics, University of Tokyo, \\
  Bunkyo-ku, Tokyo, 113--0033, Japan
}
\abstract{
Several basic results are reviewed on purpose to construct 
the quantum field theory including gravity, based on 
physical assumptions as few as possible. Up to now, 
the work by Steven Weinberg probably suits this purpose the most.
Motivated by these results we focus on the fact that the 
dimension of an operator is not unique unless the operand is 
identified. This leads to the classification 
of possible singularities for the relativistic 
Schr\"{o}dinger equation.}
\section{Introduction}
\hspace{4.2mm} 
I first thank you for giving me the chance to overview in section 2 
a thesis titled `Quantum Gravity with Minimal Assumptions'\cite{Mu2 }.
This is mainly the review of quantum gravity from particle point of view.
The purpose is to construct the quantum field 
theory including gravity, based on 
physical assumptions as few as possible.
This consists of 5 subjects, the last of which  
is an original consideration on the relation between 
essential singularity and renormalization.
This subject is summarized in section 3-5, but please read 
a preprint\cite{Mu1 } for more details.
\section{Overview}
\hspace{4.2mm} The first subject, and probably suits this purpose the most is the work by Steven Weinberg, in which he derived the Einstein equation from the Lorentz invariance of the S-matrix. According to his old paper\cite{Wein }, gravity is derived without assuming a curved space-time. Therefore, the general covariance and the geometric property of gravity are possibly subsidiary or mere approximations.

The second subject is that, according to an effective field theory, we
can make a prediction without knowing the underlying fundamental
theory. For example, John F. Donoghue calculated one loop quantum
corrections to the Newtonian potential explicitly, by assuming the
Einstein-Hilbert action and fluctuations around the flat metric, and by
making use of the result of 't Hooft and Veltman. The potential naturally 
contains the classical corrections by general relativity\cite{Dono }. 

As the third subject, we review what will happen if we loosen the
assumption on coordinates in the standard model that all physical
coordinates are transformed to the Minkowski space-time by a Poincar\'{e} transformation. And we review the troubles and the measures in treating gravitational field under classical approximations assuming a curved space-time\cite{Wald }. 
It is known that for the standard model of elementary particles, the anomaly cancellation condition in a curved space-time with torsion is the same as in a flat space-time\cite{Doba }.

As the fourth subject, we clarify the inevitable ambiguities of a
theory. The following works are reviewed. For example, the vacuum 
state in a curved space-time is not
unique and there exist several theories those can not be distinguished
by finite times of measurements\cite{Wald }. This is a theorem on
the ambiguity related to the problem of divergence. For another example,
a higher-derivative theory includes non-physical solutions those can not
be Taylor expanded. This can be the origin of the gauge ambiguity. If we
exclude superfluous solutions by imposing the perturbative constraint
conditions, it means a gauge fixing and the theory is reduced to local and lower-derivative\cite{Simon }. This treatment is  known to be equivalent to the treatment of a constraint system by Dirac brackets\cite{Dirac }. 

As the last subject, we consider the following problem. 
In usual dimensional counting, momentum has dimension one. But a
function $f(x)$, when differentiated $n$ times, does not always behave
like one with its power smaller by $n$. This inevitable uncertainty may
be essential in general theory of renormalization, including quantum
gravity.  As an example, we classify possible singularities of a
potential for the Schr\"{o}dinger equation, assuming that a potential
$V$ has at least one $C^2$ class eigen function. The result crucially
depends on the analytic property of the eigen function near its 0
points. 

Notice that neither super-symmetric, higher dimensional, nor grand 
unification theory is referred to. 
\section{Renormalization and Essential Singularity}
\hspace{4.2mm} For the rest of this article we are going to focus on 
the preprint titled Renormalization and Essential Singularity\cite{Mu1
}. 
We consider the relativistic Schr\"{o}dinger equation assuming a time 
independent and spherical symmetric $U(1)$ potential 
$A^\mu :=(\phi (r), 0, 0, 0)$. 
Then, the spherical part of an eigen function satisfies 
\beqn [-\frac1{r^2}\frac{d}{dr}(r^2\frac{d}{dr})+\frac{l(l+1)}{r^2}]y
& = & \frac{(E-e\phi )^2-m^2c^4}{{\hbar}^2c^2}y\nonumber \\
& =: & -V(r)y\;\;  \label{1} \eeqn
From now on, $V(r)$ defined in the R. H. S. is called a
potential if and only if there exists a $C^2$ class eigen function $y(r)$
satisfying (\ref{1}). The problem is, how singular $V(r)$ can be.

For simplicity we first treat 1 dimensional case with the angular 
momentum $l=0$. Then 
\eq \displaystyle{(\ref{1}) \iff \frac{y"}{y}=V(r)} \;\; , \en
so not $y(r)$ itself but the {\bf ratio} is important. For example, if $y(r)$ is Taylor expanded, 
the second derivative of the constant and the linear term vanish. That is,
\beqn \frac{y"}{y}=\frac{0 +0 +2cr+\cdots}{a +br+cr^2+\cdots}\;\; .
\nonumber \eeqn
Therefore, the singularity of a potential depends on whether or not $a, b=0$. 
In fact there are various kinds of singularities\cite{Comp }\cite{Pen }\cite{Ahlf }. For example, we can
replace the power of any term of a Taylor series with an arbitrary real
number $n$, or $\log r$. An infinite power is called an essential
singularity, and we can make more and more complex singularities by
finite times of operations including summations, subtractions,
multiplications, divisions, and compositions. 

The most general shape of a singularity that is closed
 in these operations is like 
\beqn f(z) & := & (1)_i +(2)_j +\cdots +(m)_k \; , \nonumber \\
(1)_i \;\; & := & \mbox{\huge (}\sum_{n\in \{ n\}_i 
}^\infty\sum_{m_1,\cdots , m_{d_i }=-\infty }^{m_{i1 }, \cdots , m_{id_i 
} }a_{inm_1\cdots m_{d_i } }z^n(-\log z)^{m_1 }(-\log (-z/\log 
z))^{m_2}\nonumber \\ &    &  \hspace{4.2 cm}\cdots (-\log (-z/(-\log  
(-z/\log\cdots z))))^{m_{d_i } } \mbox{\huge )}_i, \nonumber \\ 
(2)_{\pm j } & := & \sum_{i\in \{ i\}_j }(\pm )e^{\pm (1)_i },  
\nonumber \\ 
(3)_{\pm k } & := & \sum_{j\in \{ j\}_k }(\pm )e^{\pm (2)_j },  
\nonumber \\ 
\vdots \;\; .\label{ko24 } \eeqn 
More precise construction and the meaning of this expansion are in \cite{Mu1 }. Notice that this has several number of infinite series in one
expansion and all the terms are partially ordered in the ascending
powers of $r$. In this case, the domain of the power of $V(r)$ 
in the limit $r\to +0$ is 
\eq V(r)\to r^\nu , -2+\epsilon\lesssim\nu\; ;-1\leq\nu ,\en
where $\epsilon$ means an infinitesimal positive power
 like $-(log r)^{-1}$. 

Thus we can restrict the shape (i.e. power and sign) of a potential
 $V(r)$.  This is the short distance limit case, but we can also treat the
 long distance limit case by the change of variables and in dimension N
 there are 10 possible cases. Although precise version is in
 \cite{Mu1 }, we can see a shortcut version of the derivation of this main result in the next section. 

\section{Main Results}
\hspace{4.2mm} Here is the summary of the calculation. If we 
assume that the eigen function $y(r)$ is a $N$-dimensional spherical 
symmetric function $R(r)$ (i.e. orbital angular momentum $l=0$),  
and that $R(r)$ is  $C^2$ class, then (\ref{ko24 }) can be expanded as
\footnote{
The coefficients are all real and $b_i, c_j, d_k,\cdots$ are positive if 
exist.} 
\beqn R=a + br  & + & \sum_{n=2 }^\infty  
a_nr^n\sim +\cdots +\sum_{i<0 }(\pm )e^{-b_ir^i\sim\cdots } 
\cdots  \nonumber \\ 
 & + & \sum_{j<0 }(\pm )e^{-e^{c_jr^j\sim\cdots }. 
.. }\cdots +\sum_{k<0 }(\pm )e^{-e^{e^{d_kr^k\sim\cdots }\cdots } 
\cdots }\cdots . \label{ko27' } \eeqn 
For $a=0$ and $N\neq 1$, the behavior of $V(r)$ in the limit $r\to +0$ is 
\beqn \frac{\Delta R(r) }{R(r)} & = & 
 \frac{R''}{R}+\frac{N-1}{r}\frac{R'}{R} \nonumber \\ 
 & \to & \left \{ \begin{array}{l} +(N-1)r^{-2}\;\; (b\neq 0) \\ 
+n(n+N-2)r^{-2 }\;\; (b=0\; \mbox{and} \;  ^\exists a_n\neq 0) \\ 
+(-ib_i)^2r^{2i\pm 2\epsilon -2 }\;\; (b= ^\forall a_n=0\; \mbox{and} \;  ^\exists  
b_i>0) \\ 
+\infty\;\; (b= ^\forall a_n= ^\forall b_i=0\; \mbox{and} \;  ^\exists  
c_j\; \mbox{or} \; d_k\;  
\mbox{or} \cdots >0)\end{array} \right. .\label{ko30 } \eeqn 
 
We can extend the results to $r\to +\infty$ case as follows. If we 
 change the variable $r$ to $z:=\frac{1}{r}$ and assume that $R(z)$ is  
$C^2$ class (expanded like above)  
(\ref{ko30 }) is clearly replaced by 
\beqn \frac{\Delta R(r) }{R(r)} & = & 
 \frac{1}{R(z)}\left \{{\frac{dz}{dr}\frac{d}{dz} 
\left ({\frac{dz}{dr}\frac{dR(z)}{dz}}\right ) 
+(N-1)z\frac{dz}{dr}\frac{dR(z)}{dz}}\right \} \nonumber \\ 
 & = & z^4\frac{R''(z)}{R(z)}-z^3(N-3)\frac{R'(z)}{R(z)} \nonumber \\ 
 & \to & \left \{ \begin{array}{l} (3-N)\frac{b}{a}z^{3}\;\; (a\neq 0\; 
  \mbox{and} \; b\neq 0\; \mbox{and} \;  N\neq 3) \\ 
(n-N+2)n\frac{a_n}{a}z^{n+2}\;\; (a\neq 0\;  \mbox{and} \; b=0\; \mbox{and} \;  
^\exists a_n\neq 0\; \mbox{and} \;  N\neq 3) \\ 
(n-1)n\frac{a_n}{a}z^{n+2}\;\; (a\neq 0\; \mbox{and} \; ^\exists a_n\neq 0\;  
\mbox{and} \;  N=3) \\ 
(\pm )\; 0\;\; (a\neq 0\; \mbox{and} \; b=^\forall a_n=0 \; \mbox{and} \; 
 ^\exists b_i\; \mbox{or} \; c_j\; \mbox{or} \; d_k\; \mbox{or} \cdots >0) \\ 
(3-N)z^{2}\;\; (a=0\; \mbox{and} \; b\neq 0\; \mbox{and} \;  N\neq 3) \\ 
(n-1)n\frac{a_n}{b}z^{n+1}\;\; (a=0\;  \mbox{and} \; b\neq 0\; \mbox{and} \;  
^\exists a_n\neq 0\; \mbox{and} \;  N=3) \\ 
(\pm )\; 0\;\; (a=0\;  \mbox{and} \; b\neq 0\; \mbox{and} \; ^\forall a_n=0\;  
\mbox{and} \; ^\exists b_i\; \mbox{or} \; c_j\; \mbox{or} \; d_k\; \mbox{or} \cdots  
>0\; \mbox{and} \;  N=3) \\ 
(n-N+2)nz^{2}\;\; (a=b=0\; \mbox{and} \; ^\exists a_n\neq 0) \\ 
+(-ib_i)^2z^{2i\pm 2\epsilon +2 }\;\; (a=b= ^\forall a_n=0\; \mbox{and} \;  ^\exists  
b_i>0) \\ 
+\infty\;\; (a=b= ^\forall a_n= ^\forall b_i=0\; \mbox{and} \;  ^\exists  
c_j\; \mbox{or} \; d_k\;  
\mbox{or} \cdots >0)\end{array} \right. .\label{ko31 } \eeqn 
Noting that $2\leq n$ and $i<0$, we conclude that the power of potential 
 $V(r)\to r^\nu$ as $r\to \infty$ is $\nu\leq -3\; ; \;\;-2-\epsilon\lesssim\nu$. 
There is no reason to assume that $R(z)$ is $C^2$ class, but more 
 natural normalizability condition that $R(r)$ is a $L^2$ function leads  
 to small modification $a=b=0$ and $N<2n$ (instead of $2\leq n$) in 
 (\ref{ko27' }) and (\ref{ko31 }). Notice that (\ref{ko30 }) for more 
general case of $N, a$ can be obtained from (\ref{ko31 }) by the trivial replacement $N\to 4-N$ and $z\to r$ with its power smaller by $4$. Furthermore, 
above results show that for a physical dimension $N=1, 2, 3$, 
the sign of a potential $V$ must be positive for
$\nu\lesssim -2+\epsilon\;\; (r\to 0)$ and 
$-2-\epsilon\lesssim\nu\;\; (r\to\infty)$, but can be negative for other cases.

We considered here an example of the real scalar field, and the fermion field 
equation is of course another story. As a future work, I'd like to apply the 
result to the general theory of renormalization, and renormalons 
appearing in the perturbative QCD. 
\section{Discussions}
\hspace{4.2mm}{\it A potential with $C^2$ class eigen function} is only an assumption. 
The definition of a potential here is local, and valid only in this paper. 
But my conjecture is that the analyticity of an eigen function is such a 
property that cannot be distinguished by finite times of measurements. 
Therefore we assumed that a physical eigen function is $C^2$ class. 
This is also only a local constraint and very weak self consistent condition. 
Therefore it is not a sufficient condition and normalizability is another 
problem, particularly, in the long distance limit case.

{\it The motivations} to introduce such a condition are as follows. The first is that we want to clarify the inevitable ambiguity of a theory. The second is 
that there are subtle physical problems: which is the more fundamental,
a matter field or a potential? Can a potential be measured in the
absence of a matter field? It is a future work to clarify the relation between all the different facts that a quantity is realistic, physically measurable, reduced inevitably from other properties, and calculable.
\section*{Acknowledgements} 
I am grateful to Izumi Tsutsui and Toyohiro Tsurumaru for  
useful discussions. This article is partially based on 
some implications given by Tsutomu Kambe and Kazuo Fujikawa. 
I also appreciate Tsutomu Yanagida and Ken-ichi Izawa, and 
all related people.

\end{document}